# Topology Optimization under Thermo-Elastic Buckling

Shiguang Deng, Krishnan Suresh *

Mechanical Engineering, University of Wisconsin, Madison, USA

**Abstract**

The focus of this paper is on topology optimization of continuum structures subject to thermally induced buckling. Popular strategies for solving such problems include Solid Isotropic Material with Penalization (SIMP) and Rational Approximation of Material Properties (RAMP). Both methods rely on material parameterization, and can sometimes exhibit pseudo buckling modes in regions with low pseudo-densities.

Here we consider a level-set approach that relies on the concept of topological sensitivity. Topological sensitivity analysis for thermo-elastic buckling is carried out via direct and adjoint formulations. Then, an augmented Lagrangian formulation is presented that exploits these sensitivities to solve a buckling constrained problem. Numerical experiments in 3D illustrate the robustness and efficiency of the proposed method.

## 1. Introduction

Topology optimization has rapidly evolved from an academic exercise into an exciting discipline with numerous industrial applications [1], [2]. Applications include optimization of aircraft components [3], [4], spacecraft modules [5], automobiles components [6], cast components [7], compliant mechanisms [8]–[11], etc.

The focus of this paper is on topology optimization of structures subject to thermo-elastic buckling. As an illustrative example, consider the wing rib structure of a high Mach supersonic aircraft in Figure 1. During rocket boost, the aircraft is subject both to rapid acceleration and significant thermal gradients, with surface temperature as high as $1650^0 C$. Since the rib structures are welded onto the wing skins, uneven thermal heating may induce significant compressive stresses, resulting in buckling. Therefore, such structural components operating in extreme thermal environment must be designed to resist thermo-elastic buckling.

Thermo-elastic buckling poses both theoretical and computational challenges. In Section 2, popular methods for buckling-constrained topology optimization are reviewed, and the challenges are identified. In Section 3, we provide a brief overview of topological sensitivity based optimization; this is followed by the proposed method and its implementation. In Section 4, numerical experiments are presented, followed by conclusions in Section 5.

## 2. Literature review

### 2.1 buckling constrained topology optimization

Buckling typically occurs in thin-walled structures [12]. Buckling constrained topology optimization was originally studied using ground structure (truss based) approaches, while more recent methods are continuum based; the latter can be classified into the following types: Solid Isotropic Material with Penalization (SIMP), evolutionary structural optimization (ESO) and level-set. The ground structure and continuum methods are reviewed next.

#### 2.1.1 Ground structure

Ground structure approach is the classic method for optimizing the topology of truss systems. In this approach, a network of truss members is first prescribed in a design domain. A size optimization is carried out on each truss member until the cross-section areas of non-optimal trusses approach zero, and can therefore be removed [13].

However, including buckling constraint in truss optimization is non-trivial. The forces in each truss member must satisfy constraints which discontinuously depend on design variables [14]. Traditional optimizers face difficulty in solving such problems. In [14], the author argued that including slenderness constraints into buckling problems can guarantee existence of solution, and simplify the algorithm. In [15], by using a smoothing procedure to remove singularity, size optimization was made more efficient. In a recent publication [16], the author used a mixed variable formulation to linearize buckling constraint over each structural member.

#### 2.1.2 SIMP

In continuum topology optimization, the most popular method is Solid Isotropic Material with Penalization (SIMP). Its primary advantages are that it is well understood, robust and easy to implement [17]. Indeed, SIMP has been applied to a variety of topology optimization problems ranging from fluids to non-linear structural mechanics.

In thermo-elastic topology optimization, it was pointed out in [18] that the material interpolation used in SIMP exhibits zero slope at zero density, leading to robustness issues. To overcome this deficiency the Rational Approximation of Material Properties (RAMP) was developed, and its superior performance over SIMP was published in [19]. In [20], [21], a porous material penalization model was proposed for both macroscopic and microscopic material densities. It was also argued that in thermo-elasticity, porous material model with optimal microstructures perform better. In [22] a robust three-phase topology optimization technique was used to design a multi-material thermal structures with low thermal expansion and high structural stiffness.

In buckling constrained topology optimization, the appearance of pseudo buckling modes in low-density regions can pose problems. In [23], a buckling load criterion was introduced to ignore the geometric stiffness matrix of the elements whose density and principal stress were smaller than a prescribed value. In [24], the author argued such cut-off methods may abruptly change the objective function and sensitivity field, leading to

---

oscillation. Instead, the author suggested using different penalization scheme for stiffness matrix and geometric stiffness matrix. Although the author in [25] suggested it was difficult to select an appropriate penalty scheme for accurate calculation of buckling load factor, the proposed approach by [24] became a popular formulation for many researchers [26]. In a recent publication [27], a new approach to remove pseudo buckling mode was based on eigen-value shift, and pseudo mode identification.

### 2.1.3 ESO

ESO [28] is an alternate topology optimization formulation where finite elements are gradually removed based on their significance with respect to the objective function. BESO [29] addresses some of the limitations of ESO by permitting insertion of elements. In [30], a modified ESO method was proposed to maximize buckling load factor. The sensitivity of the lowest eigen-value was first derived, and the buckling eigen-value maximization was then formulated by suitably selecting the optimization criteria.

### 2.1.4 Level-set

The concept of level-set was first proposed in [31] to model the evolution of interfaces in multi-phase flows, and image segmentation problems [32]. In structural optimization, the level-set method is used to capture the evolving topology, and this leads to well-defined boundaries over which mechanical response can be accurately computed, avoiding the ambiguity associated with density-based approaches.

Specifically, in [33], the level-set method was coupled with ESO to nucleate holes, and to move boundaries based on an evolutionary stress criteria. In [34], the level-set was exploited to create a shape sensitivity based optimization framework. This was later developed to include topological derivative, and implemented in a shape sensitivity based level-set formulation [35], [36]. The initial approach to propagate level-sets was through Hamilton-Jacobi equation [34], but this was gradually replaced by mathematical programming due to higher efficiency and better constraint control [37].

In recent years, the level-set method has been extended to a variety of problems. For example, in [38], a level-set method was implemented to minimize structural compliance while maintaining fiber paths smooth, and manufacturability for steered fiber composites.

With specific reference to buckling-constrained topology optimization, in [39], a simplified buckling sensitivity field was incorporated into a level-set based framework to accelerate large-scale topology optimization process, however thermally induced buckling was not considered. It is noted the switching of critical buckling eigen-mode during optimization can cause convergence difficulty. This paper does not address mode-switching; instead, our focus is on thermally induced buckling topology optimization problems.

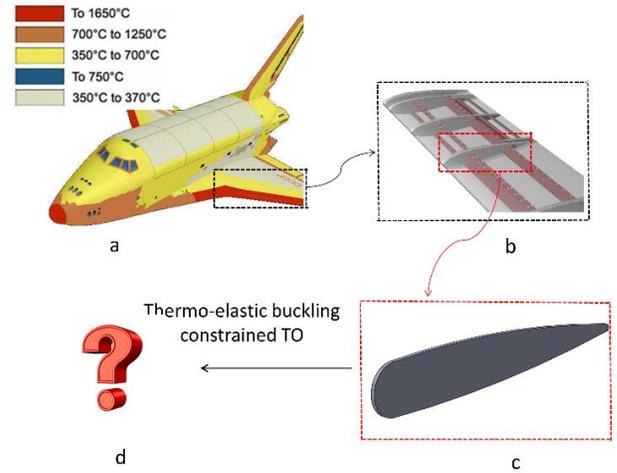

**Figure 1:** (a) Aircraft operating in high temperatures; (b) Wing design should consider thermal loads; (c) The underlying wing rib structure; (d) Optimizing the rib structure to resist thermo-elastic buckling.

### 2.2 Research gaps

From the above literature review, one can conclude that there is significant interest in solving buckling constrained topology optimization problems, with the following research gaps:

- Although thermal buckling is of significant importance, there has been very little research on thermally induced buckling constrained topology optimization.
- Computing buckling sensitivity is expensive, and efficient approaches are needed.
- Prior researchers have validated their algorithms on simple, typically 2D, examples. Extensions to large scale 3D problems is desirable.

In this paper, the objective is to minimize the volume fraction, subject to several constraints, including thermo-elastic compliance and thermo-elastic buckling load factor. Specifically, we consider the topological level-set method proposed in [39] for pure-elastic buckling problems, and extend this to thermo-elastic buckling through two distinct approaches: direct and adjoint methods. In the topological level-set method, instead of relying on the Hamilton-Jacobi equations for level-set propagation [40], fixed-point iteration is exploited to advance the topology [41].

## 3. Problem formulation and algorithm

### 3.1 Optimization problem formulation

A generic thermo-elastic bucking and compliance constrained topology optimization problem may be posed as:

$$\underset{\Omega \subset D}{Min} |\Omega|$$
$$J \leq a_1 J_0$$
$$P \geq a_2 P_0$$
subject to
$$\mathbf{K}_t \mathbf{t} = \mathbf{q}$$
$$\mathbf{Kd} = \mathbf{f} = \mathbf{f}_m + \mathbf{f}_{th}$$
(1)

where:

$\Omega$ : Topology to be computed
D : Design domain
$\mathbf{t}$ : Temperature field
$\mathbf{d}$ : Displacement field
$\mathbf{K}_t$ : Thermal stiffness matrix
$\mathbf{K}$ : Structural stiffness matrix
$\mathbf{q}$ : Heat flux
$\mathbf{f}$ : Total external load (2)
$\mathbf{f}_{st}$ : Mechanical load
$\mathbf{f}_{th}$ : Thermal load
$J$ : Compliance
$J_0$ : Initial compliance
$P$ : Buckling load factor
$P_0$ : Initial buckling load factor
$a_1, a_2$ : User defined scalars

In other words, the objective is to find the optimal topology with minimal volume within the design domain (D) while satisfying prescribed compliance and buckling constraints. During the optimization process, the displacement and temperature fields are calculated from thermo-elastic finite element analysis.

### 3.2 Thermo-elastic FEA

For completeness, we summarize finite element formulations of (weakly-coupled) thermo-elastic problems; recall that such problems reduce to solving two linear equations [12]:

$$\mathbf{K}_t \mathbf{t} = \mathbf{q} \tag{3}$$

$$\mathbf{Kd} = \mathbf{f} = \mathbf{f}_{st} + \mathbf{f}_{th} \tag{4}$$

The elemental thermal load vector in Equation (4) is formed via [12], [42]:

$$\mathbf{f}_e^{th} = \int_{\Omega_e} \mathbf{B}^T \mathbf{D} \boldsymbol{\varepsilon}_e^{th} d\Omega \tag{5}$$

$$\boldsymbol{\varepsilon}_e^{th} = \alpha (t_e - t_0) \boldsymbol{\Phi}^T \tag{6}$$

where:

$\mathbf{f}_e^{th}$ : Element thermal load vector
$\Omega_e$ : Element domain
$\mathbf{B}$ : Element strain-displacement matrix
$\mathbf{D}$ : Element elasticity matrix
$\boldsymbol{\varepsilon}_e^{th}$ : Element thermal strain vector (7)
$\alpha$ : Thermal expansion coefficient
$t_e$ : Element temperature
$t_0$ : Reference temperature
$\boldsymbol{\Phi}$ : [1 1 1 0 0 0] in 3D; [1 1 0] in 2D

The stresses are obtained by subtracting the thermal strain from the total strain, and multiplying the resulting strain by the material tensor:

$$\boldsymbol{\sigma}_e = \mathbf{DBd}_e - \mathbf{D}\boldsymbol{\varepsilon}_e^{th} \tag{8}$$

Further details may be found, for example, in [43]. The compliance for a thermo-elastic system is defined as:

$$J = (\mathbf{f} + \mathbf{f}^{th})^T \mathbf{d} = \mathbf{d}^T \mathbf{Kd} \tag{9}$$

However, if there is no structural load ($\mathbf{f} = 0$ in Equation (9)), minimizing compliance is questionable since in the absence of an external load, the best structure is no structure. Therefore, in this paper, we *consistently assume there exists a structural load to prevent the ill-posed problem.*

We also assume that the temperature within the design domain is spatially uniform, i.e., solving Equation (3) becomes unnecessary. If the temperature fields are dependent on the design [44], the calculation of topological sensitivity must necessarily involve solving Equation (3).

Finally, observe that Equation (4) represents a weakly-coupled problem where the thermal field influences the displacements, but not the reverse. Strongly-coupled thermo-elastic problems are beyond the scope of this paper.

### 3.3 Buckling FEA

The linear buckling load factor can be calculated from the well-known formulation [12]:

$$(\mathbf{K} + \lambda \mathbf{K}_\sigma)\boldsymbol{\upsilon} = \mathbf{0} \tag{10}$$

where

$\mathbf{K}_\sigma$ : Global geometric stiffness matrix
$\lambda$ : Linear buckling load factor
$\boldsymbol{\upsilon}$ : Buckling mode vector

In Equation (10), the global geometric stiffness matrix is defined via the assembly:

$$\mathbf{K}_\sigma = \sum_{e=1}^{N} [\mathbf{k}_\sigma]_e \tag{11}$$

where N is the number of finite elements and the elemental geometric stiffness matrix are defined as:

$$[\mathbf{k}_\sigma]_e = \int_{\Omega_e} \mathbf{G}^T \mathbf{S} \mathbf{G} dv \tag{12}$$

where $\mathbf{G}$ is obtained from shape functions by appropriate differentiation and reordering [12]. The matrix $\mathbf{S}$ is defined as:

$$\mathbf{S} = \begin{bmatrix} \mathbf{s} & \mathbf{0} & \mathbf{0} \\ \mathbf{0} & \mathbf{s} & \mathbf{0} \\ \mathbf{0} & \mathbf{0} & \mathbf{s} \end{bmatrix} \tag{13}$$

where

$$\mathbf{s} = \begin{bmatrix} \sigma_x & \tau_{xy} & \tau_{xz} \\ \tau_{xy} & \sigma_y & \tau_{yz} \\ \tau_{xz} & \tau_{yz} & \sigma_z \end{bmatrix} \tag{14}$$

In addition, the stress in an element is defined as:

$$[\boldsymbol{\sigma}]_e = [\sigma_x, \sigma_y, \sigma_z, \tau_{xy}, \tau_{xz}, \tau_{yz}]^T \tag{15}$$

It is clear that, in Equation (10), the geometric stiffness matrix $\mathbf{K}_\sigma$ is a function of stress ($\boldsymbol{\sigma}$) which in turn depends on the topology, while the stiffness matrix ($\mathbf{K}$), buckling load factor ($\lambda$) and buckling mode vector ($\boldsymbol{\upsilon}$) are explicitly dependent on the topology. It is also noted since the temperature field is uniformly elevated to a prescribed value, and the temperature field ($\mathbf{t}$) is not dependent on the topology.

A practical challenge that arises in solving large-scale topology optimization is the computational costs rising from the underlying FEA. To address the computational cost, we rely here on the assembly-free deflated conjugate gradient (AF-DCG) method proposed in [45]. The assembly free method rests on the observation that the computational bottle-neck in modern architecture is memory access [46]. The AF-DCG computes the preconditioner and the solution to the underlying linear system in an assembly-free manner, significantly reducing memory bandwidth, and therefore speeding up FEA.

### 3.4 Buckling sensitivity analysis

We now focus our attention on computing sensitivities that are essential for topology optimization. Let Q be any quantity of interest in an optimization problem. The sensitivity of Q with respect to a topological design variable $\mathbf{x}$ is denoted by:

$$Q' \equiv \frac{\partial Q}{\partial \mathbf{x}} \tag{16}$$

The derivatives of the global stiffness matrix and geometric stiffness matrix will be denoted by:

$$\mathbf{K}' \equiv \frac{\partial \mathbf{K}}{\partial \mathbf{x}} \tag{17}$$

$$\mathbf{K}_\sigma' \equiv \frac{\partial \mathbf{K}_\sigma}{\partial \boldsymbol{\sigma}} \boldsymbol{\sigma}' \tag{18}$$

In this section, two approaches are used to calculate the sensitivity, specifically for the linear buckling load factor $\lambda$.

#### 3.4.1 Direct method

Multiplying the buckling mode vector ($\boldsymbol{\upsilon}^T$) on both sides of Equation (10), and taking the derivative with respect to design variable, we have:

$$2\boldsymbol{\upsilon}'^T(\mathbf{K} + \lambda \mathbf{K}_\sigma)\boldsymbol{\upsilon} + \boldsymbol{\upsilon}^T(\mathbf{K}' + \lambda \mathbf{K}_\sigma' + \lambda' \mathbf{K}_\sigma)\boldsymbol{\upsilon} = 0 \tag{19}$$

Due to Equation (10), the first term in Equation (19) vanishes. Reordering terms in Equation (19), we have the sensitivity of the linear buckling load factor as:

$$\lambda' = -\frac{\boldsymbol{\upsilon}^T(\mathbf{K}' + \lambda \mathbf{K}_\sigma')\boldsymbol{\upsilon}}{\boldsymbol{\upsilon}^T \mathbf{K}_\sigma \boldsymbol{\upsilon}} \tag{20}$$

A simple method to calculate $\mathbf{K}_\sigma'$ is to use finite difference. Obviously, this method is too expensive and potentially inaccurate. Instead, we consider the following direct approach first.

Observe that Equation (18) can be written as the summation over all finite elements:

$$\frac{\partial \mathbf{K}_\sigma}{\partial \boldsymbol{\sigma}} \boldsymbol{\sigma}' = \sum_{j=1}^{N} \left( \frac{\partial \mathbf{K}_\sigma}{\partial \boldsymbol{\sigma}_j} \boldsymbol{\sigma}_j' \right) \tag{21}$$

where $N$ is the number of all finite elements, i.e., the sensitivity of the global geometric stiffness matrix is the summation of sensitivities over all elements.

For a specific element 'j' in Equation (21), one can further expand the sensitivities over the six stress components:

$$\frac{\partial \mathbf{K}_\sigma}{\partial \boldsymbol{\sigma}_j} \boldsymbol{\sigma}_j' = \sum_{k=1}^{6} \left( \frac{\partial \mathbf{K}_\sigma}{\partial \sigma_{jk}} \sigma_{jk}' \right) \tag{22}$$

Further, by definition:

$$\frac{\partial \mathbf{K}_\sigma}{\partial \sigma_{jk}} = \frac{\partial \left( \int_{\Omega_j} \mathbf{G}^T \mathbf{S} \mathbf{G} dv + \sum_{i=1, i \neq j}^{N} \left( \int_{\Omega_i} \mathbf{G}^T \mathbf{S} \mathbf{G} dv \right) \right)}{\partial \sigma_{jk}} \tag{23}$$

Since the geometric stiffness matrices in elements other than 'j' are not explicitly dependent on the stress in j-element, the second term in numerator of Equation (23) drops out:

$$\frac{\partial \mathbf{K}_\sigma}{\partial \sigma_{jk}} = \frac{\partial \int_{\Omega_j} \mathbf{G}^T \mathbf{S} \mathbf{G} dv}{\partial \sigma_{jk}} = \int_{\Omega_j} \mathbf{G}^T \frac{\partial \mathbf{S}}{\partial \sigma_k} \mathbf{G} dv \tag{24}$$

where

$$\frac{\partial \mathbf{S}}{\partial \sigma_k} = \begin{bmatrix} \frac{\partial \mathbf{s}}{\partial \sigma_k} & 0 & 0 \\ 0 & \frac{\partial \mathbf{s}}{\partial \sigma_k} & 0 \\ 0 & 0 & \frac{\partial \mathbf{s}}{\partial \sigma_k} \end{bmatrix} \tag{25}$$

For the six stress components in Equation (15), it is easy to calculate the contributions in Equation (25). For example, when $k=1$, we have:

$$\frac{\partial \mathbf{s}}{\partial \sigma_1} = \begin{bmatrix} 1 & 0 & 0 \\ 0 & 0 & 0 \\ 0 & 0 & 0 \end{bmatrix} \tag{26}$$

Next, the term ($\sigma_{jk}'$) in Equation (22) can be derived as follows. Rewrite Equation (8) for the j-element:

$$\boldsymbol{\sigma}_j = \mathbf{D} \mathbf{B} \mathbf{d}_j - \mathbf{D} \boldsymbol{\varepsilon}_j^{th} \tag{27}$$

where the elemental thermal strain ($\boldsymbol{\varepsilon}_j^{th}$) can be calculated in Equation (6). Since we have assumed a uniform temperature increase, this is independent of design variable. Taking the derivative of each term in Equation (27), we have:

$$\boldsymbol{\sigma}_j' = \mathbf{D}' \mathbf{B} \mathbf{d}_j + \mathbf{D} \mathbf{B} \mathbf{d}_j' \tag{28}$$

We can calculate the term ($\mathbf{d}_j'$) in Equation (28) in the following manner. Taking derivative of the static equilibrium equation in Equation (4):

$$\mathbf{K}' \mathbf{d} + \mathbf{K} \mathbf{d}' = \mathbf{f}_{th}' \tag{29}$$

where the structural force is assumed to be independent of design variable. Reordering terms, we have

$$\mathbf{d}' = \mathbf{K}^{-1}(\mathbf{f}_{th}' - \mathbf{K}'\mathbf{d}) \tag{30}$$

The elemental displacement sensitivity in j-element ($\mathbf{d}_j'$) can be directly obtained from Equation (30).

In summary, the direct method proceeds as follows: (1) the derivative of the global geometric stiffness matrix is computed with respect to each stress component for every element using Equation (24); (2) the derivative of the global stress vector is computed using Equation (28); (3) the two results are combined using Equation (21) and Equation (22); and (4) finally Equation (20) is used to arrive at the final sensitivity of the linear buckling load factor $\lambda$.

The direct method is easy to derive and implement. However, it will be demonstrated in the numerical experiments that it is computationally inefficient for the following reason: calculating $\lambda'$ in Equation (30) requires solving a global problem for each element. This is impractical even for simple finite element models.

### 3.4.2 Adjoint method

An alternative and efficient way is by using adjoint variables and constraints. By carefully selecting the adjoint variables, the computationally expensive terms can be eliminated. This was first proposed in [47] for structurally inducted buckling problems. Here, we consider its generalization to thermo-elastic problems.

Multiplying buckling mode vector ($\upsilon^T$) on both sides of Equation (10) and augmenting this with the two constraints multiplied by two suitable adjoint variables ($\mu$) and ($\mathbf{w}$), we have:

$$\upsilon^T(\mathbf{K}+\lambda\mathbf{K}_\sigma)\upsilon + \mu^T[\sigma - \mathbf{Y}\mathbf{d}+\mathbf{Z}\varepsilon_{th}] + \mathbf{w}^T(\mathbf{f}-\mathbf{K}\mathbf{d}) = 0 \quad (31)$$

In the above equation, the matrix ($\mathbf{Y}$) and ($\mathbf{Z}$) relate displacement and thermal strains to stresses, respectively.

$$\mathbf{Y} = \sum_{j=1}^{N}[\mathbf{DB}]_j \quad (32)$$

$$\mathbf{Z} = \sum_{j=1}^{N}[\mathbf{D}]_j \quad (33)$$

In Equation (31), the adjoint $\mu$ links the stresses to deformation, and the adjoint $\mathbf{w}$ links the deformation to external load. Then taking the derivative of Equation (31) and simplifying terms, we get:

$$\upsilon^T(\mathbf{K}'+\lambda\frac{\partial \mathbf{K}_\sigma}{\partial \sigma}\sigma'+\lambda'\mathbf{K}_\sigma)\upsilon + \mu^T(\sigma' - \mathbf{Y}'\mathbf{d}-\mathbf{Y}\mathbf{d}'+\mathbf{Z}'\varepsilon_{th}) + \mathbf{w}^T(\mathbf{f}'-\mathbf{K}'\mathbf{d}-\mathbf{K}\mathbf{d}') = 0 \quad (34)$$

The first adjoint ($\mu$) is chosen such that the terms with ($\sigma'$) can be dropped from Equation (34):

$$\lambda\upsilon^T\frac{\partial \mathbf{K}_\sigma}{\partial \sigma}\sigma'\upsilon + \mu^T\sigma' = 0 \quad (35)$$

After factoring and rearranging terms, we have:

$$\mu^T = -\lambda\upsilon^T\frac{\partial \mathbf{K}_\sigma}{\partial \sigma}\upsilon \quad (36)$$

where the term $\frac{\partial \mathbf{K}_\sigma}{\partial \sigma}$ is the assembly of all elemental sensitivities, each containing six components.

$$\frac{\partial \mathbf{K}_\sigma}{\partial \sigma} = \sum_{j=1}^{N}\sum_{k=1}^{6}\frac{\partial \mathbf{K}_\sigma}{\partial \sigma_{jk}} \quad (37)$$

Equation (34) simplifies to:

$$\upsilon^T(\mathbf{K}'+\lambda'\mathbf{K}_\sigma)\upsilon + \mu^T(-\mathbf{Y}'\mathbf{d}-\mathbf{Y}\mathbf{d}'+\mathbf{Z}'\varepsilon_{th}) + \mathbf{w}^T(\mathbf{f}'-\mathbf{K}'\mathbf{d}-\mathbf{K}\mathbf{d}') = 0 \quad (38)$$

The second adjoint $\mathbf{w}$ is chosen such that the terms containing $\mathbf{d}'$ can be cancelled out:

$$\mu^T\mathbf{Y}\mathbf{d}' + \mathbf{w}^T\mathbf{K}\mathbf{d}' = 0 \quad (39)$$

After rearranging terms, we have:

$$\mathbf{w}^T = -\mu^T\mathbf{Y}\mathbf{K}^{-1} \quad (40)$$

Therefore, the sensitivity of the buckling load factor in Equation (38) can be expressed as:

$$\lambda' = -\frac{1}{\upsilon^T\mathbf{K}_\sigma\upsilon}(\upsilon^T\mathbf{K}'\upsilon + \mu^T\mathbf{Z}'\varepsilon_{th} - \mu^T\mathbf{Y}'\mathbf{d}+\mathbf{w}^T\mathbf{f}'-\mathbf{w}^T\mathbf{K}'\mathbf{d}) \quad (41)$$

In summary, the adjoint method proceeds as follows: (1) augment two adjoint terms into the original buckling expression as in Equation (31); (2) calculate the adjoints such that $\sigma'$ in Equation (36) and $\mathbf{d}'$ in Equation (39) drop out; (3) reorder the sensitivity expression as in Equation (41) to calculate $\lambda'$. Since the computation process does not involve the stiffness matrix inverse operation, the adjoint method is more efficient.

The last step is to compute the sensitivity of the global matrices in Equation (20) and Equation (41), i.e., $\mathbf{K}'$, $\mathbf{K}'_\sigma$, $\mathbf{Y}'$ and $\mathbf{Z}'$. If pseudo-density parameterization is used (as in SIMP or RAMP), then the sensitivities can be computed via their respective material interpolation scheme [48].

One disadvantage of SIMP is the introduction of localized artificial buckling modes in low pseudo-density regions [26]. In [23], the stress stiffness matrix associated with low density elements was completely neglected during stress stiffness matrix calculation. In [49], a differentiable version of interpolation schemes was proposed where the lower bound of pseudo-density was carefully selected to avoid artificial modes.

In this paper, the sensitivities are computed by evaluating the discrete topological sensitivity at the center of each element, thus avoiding the challenges with low density elements. In other words, the sensitivities are defined as follows [50]:

$$\mathbf{K}' \equiv [\mathbf{K}_e] \quad (42)$$

$$\mathbf{Y}' \equiv [\mathbf{DB}] \quad (43)$$

$$\mathbf{Z}' \equiv [\mathbf{D}] \quad (44)$$

It should also be noted that the buckling topological sensitivity field in Equation (41) are non-monotonic which means the sensitivity can take either a positive or a negative value during optimization. This non-monotonic behavior can pose challenges for traditional monotonic approximation methods. Such numerical challenge can be avoided with non-monotonous approximation methods like globally convergent version of MMA (GCMMA) and gradient-based MMA (GBMMA). In this paper, we employ topological sensitivity based level-set method that is proven to be robust for solving the non-monotonous problem, as illustrated later through numerical experiments.

We must emphasize that the above sensitivity analysis approach assumes that the buckling mode is unique. In other words, in situations where structures have multiple buckling modes, the proposed method does not guarantee local minima; readers are referred to [51] for more details. Therefore, buckling mode switching case is beyond the scope of this paper and will be studied in future research.

### 3.5 Augmented Lagrangian method

Given the expressions for sensitivities, we now consider solving the topology optimization problem in Equation (1). This is a special case of generic constrained optimization problem:

$$\underset{\mathbf{x}}{Min}\, f(\mathbf{d},\Omega) \qquad (45)$$
$$g_i(\mathbf{d},\Omega) \leq 0$$

A popular method for solving such constrained optimization method is the augmented Lagrangian method [52], where the constraints are absorbed into the objective function through:

$$L(\mathbf{d},\Omega;\gamma_i,\mu_i) \equiv f(\mathbf{d},\Omega) + \sum_{i=1}^{m} \bar{L}_i(\mathbf{d},\Omega;\gamma_i,\mu_i) \qquad (46)$$

where

$$\bar{L}_i(\mathbf{d},\Omega;\gamma_i,\mu_i) = \begin{cases} \mu_i g_i + \frac{1}{2}\gamma_i (g_i)^2 & \mu_i + \gamma_i g_i > 0 \\ \frac{1}{2}\mu_i^2 / \gamma_i & \mu_i + \gamma_i g_i \leq 0 \end{cases} \qquad (47)$$

$L$: Augmented Lagrangian
$\bar{L}_i$: Auxiliary Lagrangian
$\mu_i$: Lagrangian multipliers  (48)
$\gamma_i$: Penalty parameters
$m$: Number of constraints

Observe that the gradient of augmented Lagrangian is given by:

$$L' = f' + \sum_{i=1}^{m} \bar{L}_i' \qquad (49)$$

where

$$\bar{L}_i' = \begin{cases} \mu_i + \gamma_i g_i \; g_i' & \mu_i + \gamma_i g_i > 0 \\ 0 & \mu_i + \gamma_i g_i \leq 0 \end{cases} \qquad (50)$$

For the topology optimization problem posed in Equation (1), the objective is the volume, and therefore the topological sensitivity is given by:

$$f' = -1 \qquad (51)$$

For the constraint functions, the buckling sensitivity can be computed by Equation (20) and (41) while the sensitivity for compliance can be found in [50].

In the augmented Lagrangian method, the Lagrangian multipliers and penalty parameters in Equation (47) are initialized as follows:

$$\mu_i^0 = 1 \,,\; \gamma_i^0 = 10 . \qquad (52)$$

Then the augmented Lagrangian is minimized using, for example, conjugate gradient method. In every iteration, the multipliers are updated as follows:

$$\mu_i^{k+1} = \max\{\mu_i^k + \gamma_i g_i(\hat{\mathbf{x}}^k), 0\}, i = 1,2,3,...,m \qquad (53)$$

where the $\hat{\mathbf{x}}^k$ is the local minimum at the current $k$ iteration. The penalty parameters are updated via:

$$\gamma_i^{k+1} = \begin{cases} \gamma_i^k & \min(g_i^{k+1},0) \leq \varsigma \min(g_i^k,0) \\ \max(\eta \gamma_i^k, k^2) & \min(g_i^{k+1},0) > \varsigma \min(g_i^k,0) \end{cases} \qquad (54)$$

where $0 < \varsigma < 1$ and $\eta > 0$; typically, $\varsigma = 0.25$ and $\eta = 10$ [52].

### 3.6 Proposed method

Piecing it all together, the proposed method for thermo-elastic topology optimization (TO) combines the topological sensitivities and the augmented Lagrangian method; the algorithm proceeds as follows:

1. The optimization starts at a volume fraction of 1.0, i.e., the 'current volume fraction' v is set to 1.0, and the initial 'volume decrement' $\Delta$v is set to 0.025. The Lagrangian multipliers and parameters are initialized per Equation (52).
2. The linear static thermo-structural FEA problem in Equation (3) is solved and the stresses are extracted at the center of each element via Equation (8).
3. The linear buckling eigen-value problem in Equation (10) is then solved by using the thermal stress calculated from Step-2. The buckling topological sensitivity field is computed at the center of each element by either the direct method per Equation (20), or the adjoint method per Equation (41).
4. Using the augmented Lagrangian formulation, the sensitivity fields of the objective function and constraints are combined using Equation (49).
5. The augmented Lagrangian is then minimized. If the topology converges, the optimization moves to the next step and volume decrement is enlarged by 10%, else the volume decrement is reduced by half, and the optimization returns to Step-2. Our volume decrement algorithm is dynamic and adaptive to various optimization stages: a larger volume decrement value is used in early stage to reduce overall computational time while a much smaller value is utilized when optimization requires smaller step size for convergence. The iterations are repeated until the final volume fraction is reached or any of constraints is violated.
6. If the current volume fraction is smaller than the target volume fraction (v < v$_{\text{target}}$), the algorithm exits. Else, the volume is further reduced, and the optimization returns to Step-2. The iterations are repeated until the final volume fraction is reached or any of constraints is violated.

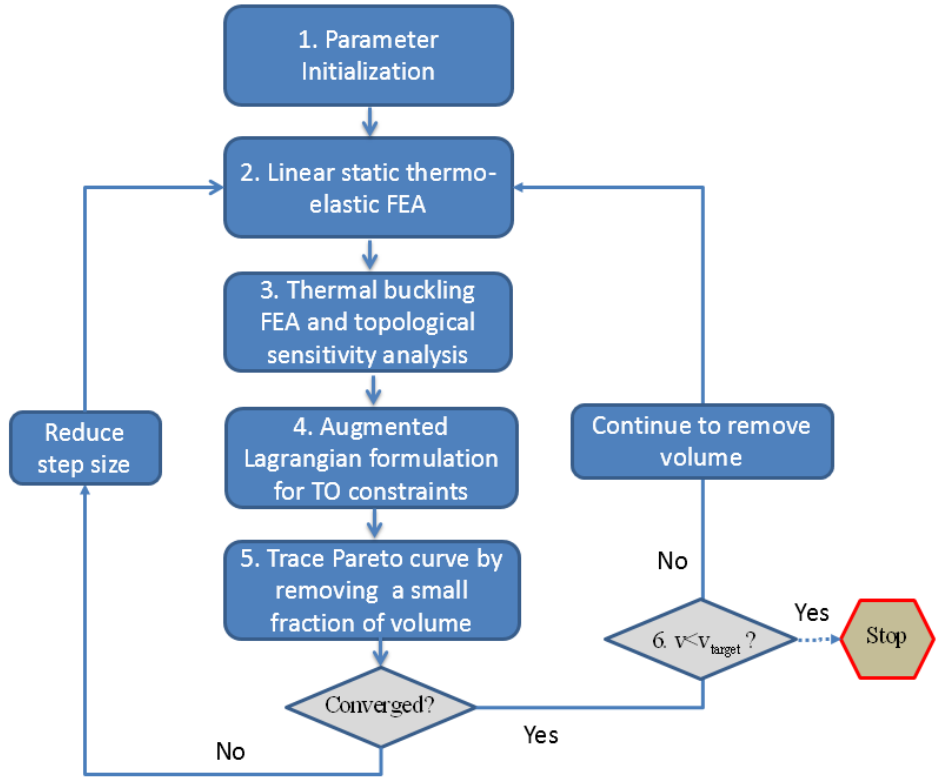

**Figure 2:** An overview of the proposed algorithm.

## 4. Numerical experiments

In this Section, we demonstrate the proposed method through numerical experiments. The default parameters are as follows:

- A thermal load is applied by increasing the temperature uniformly by $\Delta t$ with respect to a reference temperature of $t_0 = 25°C$ (the reference temperature is only relevant for determining the appropriate material properties).
- Hexahedral (8-noded) elements are used for 3D finite element analysis.
- All experiments were conducted using a C++ implementation on a Windows 10 machine, with I7-5960X, 16 GB.

**4.1 Benchmark example**

The first experiment involves the buckling of a 3D column with a width of 0.05m and a length of 0.25m, which was previously studied in [39], and is illustrated in Figure 3. The material is assumed to be steel, i.e., $E = 2e11$ Pa, $v = 0.3$ and $\alpha = 1.1e-5/°C$. As illustrated in Figure 3a, the structure is clamped at the bottom and a compressive load of $F = 1.0e5$N is applied at the center of the top edge; the structure is also subject to a homogeneous temperature elevation $\Delta t = 150°C$. Note that the 3D column will buckle out of plane as illustrated in Figure 3c.

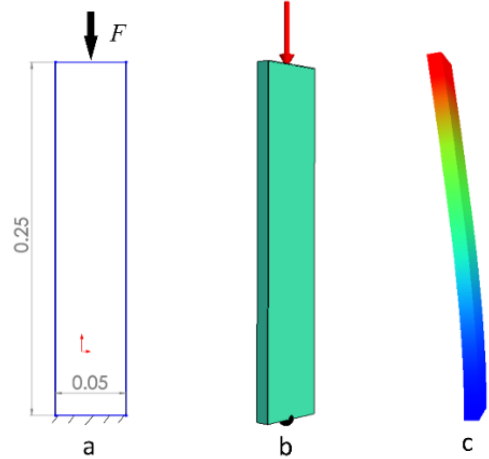

**Figure 3:** (a) A thin column structure with a thickness of 0.01m; (b) CAD model; (c) Buckling mode.

In this example, the objective is to minimize the volume fraction, while thermo-elastic compliance and thermal buckling load serve as constraints. Specifically, we search for the lightest design whose final compliance is no larger than 2.5 times its initial compliance, while the final buckling load factor is greater than or equal to 60% of its initial value. Formally, this can be expressed as:

$$\begin{aligned}&\underset{\Omega \subset D}{Min} |\Omega| \\ &J \leq 2.5 J_0 \\ &P \geq 0.6 P_0 \\ &\text{subject to} \\ &\mathbf{Kd} = \mathbf{f}_{st} + \mathbf{f}_{th} \\ &\Delta t = 150°C\end{aligned} \qquad (55)$$

### 4.1.1 Direct and adjoint methods

First, in order to compare the efficiency of direct and adjoint methods, we consider several coarse meshes, and compare the computational costs. The results are summarized in Figure 4. Observe that the proposed adjoint method is significantly faster than direct method, for reasons explained earlier. For large-scale problems, the direct method becomes impractical and will not be considered for the remainder of the paper.

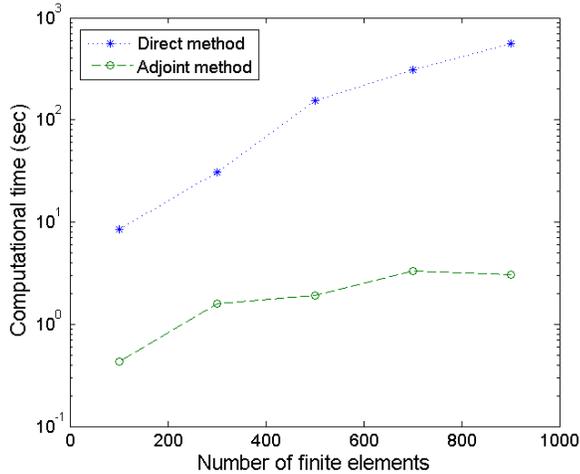

**Figure 4:** Comparison of computational time of the direct and the adjoint methods.

### 4.1.2 Importance of buckling and thermal load

Next, with the objective of studying the impact of temperature elevation and buckling constraint on topology optimization, we consider the problem in Figure 3 under three different scenarios: (a) where buckling constraint and thermal load are neglected (i.e., only compliance constraint is considered), (b) buckling constraint is neglected but thermal load is included in the compliance computation, and finally (c) where both buckling constraint and thermal load are included.

In this experiment, we use 30,000 elements (i.e., 104,832 degrees of freedom (DOF)) to discretize the design domain. The adjoint method is used due to its efficiency. The resulting topologies for the three scenarios are illustrated in Figure 5(a), Figure 5(b) and Figure 5(c), respectively. The impact of thermal load and buckling constraint are clearly observable. The observed difference in final topologies sheds light on the significance of including thermal loads in buckling constrained topology optimization. The impact of thermal load can be understood by studying Equation (4) where inclusion of the design dependent thermal load changes both the stress distribution (compared with pure elastic case) and the geometric stiffness matrix. In other words, thermal effects often lead to increased compressive loads and buckling, thereby affecting the final topology.

As expected, with additional constraints, the optimization problem terminates at a higher volume fraction (see Table 1).

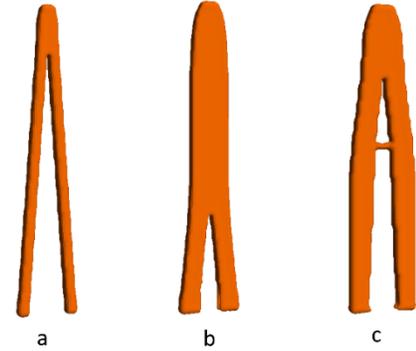

**Figure 5:** (a) Compliance constrained elastic TO; (b) Compliance constrained thermo-elastic TO; (c) Buckling and compliance constrained thermo-elastic TO.

The final volume fractions and constraints are summarized in Table 1 where the active constraints are emphasized with a 'box'. It is noted that compared to case (b), a lower volume fraction is reached in case (c) despite the additional thermal effects. One plausible reason is that, the thermal sensitivities are non-monotonic [53].

**Table 1:** Constraints and results for problem in Figure 5.

| Topology | Initial Constraints | Final Constraints | Volume & time (sec) |
|---|---|---|---|
| Figure 5(a) | $J \leq 2.5 J_0$ | $\boxed{J = 2.50 J_0}$ | v=0.33 T=67.32 |
| Figure 5(b) | $J \leq 2.5 J_0$ $P \geq 0.6 P_0$ | $J = 1.48 J_0$ $\boxed{P = 0.60 P_0}$ | v=0.59 T=145.68 |
| Figure 5(c) | $J \leq 2.5 J_0$ $P \geq 0.6 P_0$ $\Delta t = 150°C$ | $\boxed{J = 2.50 J_0}$ $P = 0.91 P_0$ $\Delta t = 150°C$ | v=0.57 T=197.03 |

For the specific case of Figure 5(c), the iteration histories with evolving topologies are illustrated in Figure 6 where the values of the compliance and buckling load factor are scaled with respect to the initial values at volume fraction of 1.0. Observe that as the volume fraction decreases, the compliance monotonously increases, while the buckling load factor generally decreases. The non-monotonic nature of buckling curve in Figure 6 was discussed earlier in Equation (23) and Equation (41).

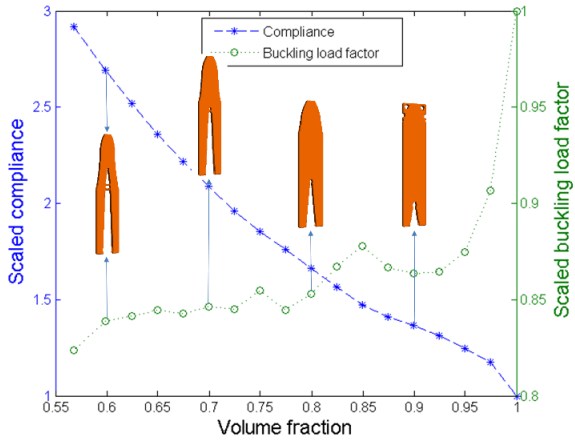

**Figure 6:** Iteration history for the topology in Figure 5(c).

**4.1.3 Mesh independence**

In this section, we study the effect of mesh size on the topology optimization results. For the case study (c) of Figure 5, we use various mesh sizes to discretize the 3D column.

The mesh sizes vary from 10,000 to 40,000 elements. The classic radial filtering technique [17] is used for smoothing topological sensitivity fields. The final topologies and corresponding volume fractions are illustrated in Figure 7 and Table 2. As one can observe, neither the final topologies nor the volume fractions are strongly dependent on the mesh size.

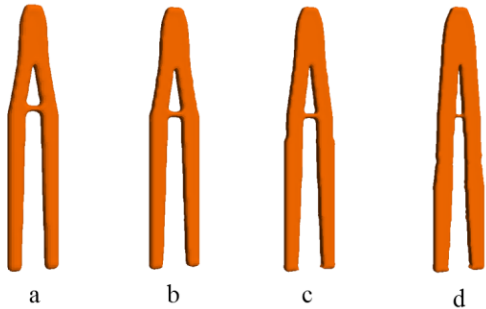

**Figure 7:** Final topologies of different mesh sizes for case study of Figure 5(c) where the mesh densities and volume fractions are listed in Table 2.

**Table 2**: Final volume fractions for mesh independence study.

| Mesh densities | Volume fraction | Final topologies |
|---|---|---|
| 10,000 | 0.58 | Figure 7(a) |
| 20,000 | 0.58 | Figure 7(b) |
| 30,000 | 0.57 | Figure 7(c) |
| 40,000 | 0.58 | Figure 7(d) |

**4.2 Industrial application: airplane wing rib structure**

The purpose of this experiment is to demonstrate the robustness of the proposed adjoint method for optimizing an airplane wing rib structure. In wing structures, to maintain wing contours in chord-wise direction, and to shorten the length of longitudinal wing stringers, ribs are used as internal supporting units as shown in Figure 8 [54].

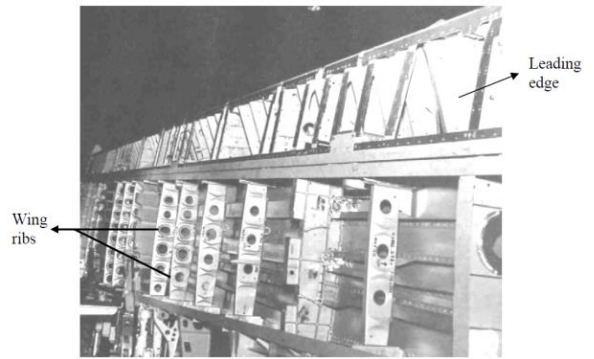

**Figure 8:** Wing rib structures with lower skin removed [54].

The rib structure consists of three distinct sections as illustrated in Figure 9: the leading-edge portion, the wing box portion and trailing edge. In the leading edge, lightening holes are often introduced for mass reduction and accessibility of wiring and pipe lines. Horizontal stiffeners are also used to prevent buckling. In the wing box portion, horizontal and vertical beads are used both to stiffen the structure and to prevent buckling. Trusses are heavily used in trailing edge portion. The rib can be welded, riveted or glutted onto wing skins. The assembly configuration can easily carry heat from hot skin (shown in **Figure 1**) to the rib structures, and the induced thermal stress may lead to buckling.

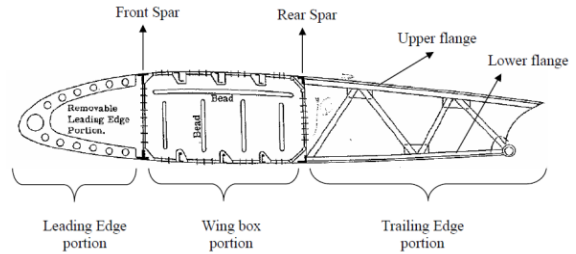

**Figure 9**: Wing rib construction [54].

Lightening holes and stiffening beads are often designed based on experience, and may not be optimal. In this section, the proposed thermo-elastic topology optimization method is used for optimizing both the leading-edge portion and wing-box portion.

**4.2.1 leading edge optimization**

During flight, the wing ribs are subject to three types of loads: (a) aerodynamic lift and drag forces, (b) concentrated forces from its connection with landing gears and fuselage, and (c) gravitational body force [54]. In this experiment, only the dominating aerodynamic forces are considered for simplicity. With speed up to 24 Mach, the lift and drag pressure on a supersonic aircraft (e.g., space shuttle) can be as high as $10^8 (\text{N/m}^2)$, while the surface temperature can be as high as $1650^\circ\text{C}$. Although thermally protected [55] the ribs underneath the skin can still reach $170^\circ\text{C} \sim 270^\circ\text{C}$ [55].

As shown in Figure 10, the leading edge is assumed to be fixed at the right edge, and loaded with a drag pressure of $146 MPa$ on the top edge and a lift pressure of $430 MPa$ at the bottom. The entire structure is subject to an increase in temperature of $\Delta t = 270°C$. The material is assumed to be titanium alloy [56] with an elastic modulus of $E = 111 GPa$, Poisson's ratio of $v = 0.33$ and thermal expansion coefficient of $\alpha = 6.0e - 6/°C$.

For FEA, 294,670 hexahedral elements are used to discretize the design domain, resulting in 972,192 DOF. The optimization problem is as follows:

$$\underset{\Omega \subset D}{Min} |\Omega|$$
$$J \leq 1.5 J_0$$
$$P \geq 0.4 P_0 \qquad (56)$$
$$\text{subject to}$$
$$\mathbf{Kd} = \mathbf{f}_{st} + \mathbf{f}_{th}$$
$$\Delta t = 270^\circ C$$

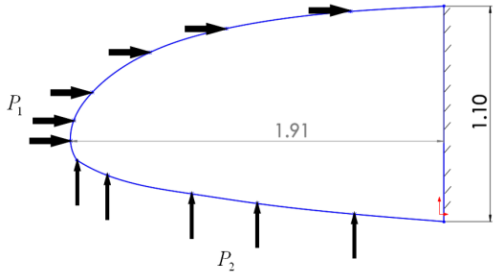

**Figure 10:** Leading-edge with a thickness of 0.1m, and applied boundary conditions (units: meters).

In words, the objective is to find the optimal topology with the minimal volume and its compliance is no more than 1.5 times the initial compliance, and its buckling load factor is no less than 40% of original value.

To illustrate the impact of thermal load, we also solve the above problem by neglecting the temperature increase. The resulting topologies for the two scenarios are illustrated in Figure 11. It can be observed that the topologies are similar, but with the thermal load, the optimization terminates at a higher volume fraction. The numerical results are summarized in Table 3.

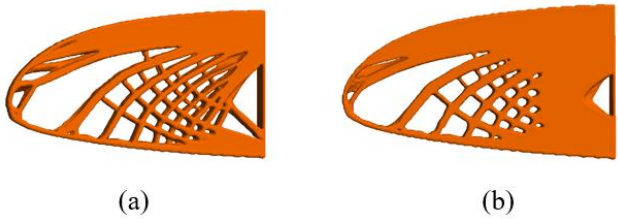

**Figure 11:** Optimal designs for the rib's leading-edge portion: (a) The design with final volume fraction of 0.52, with buckling constraint but no thermal load; (b) The design with final volume fraction of 0.7, with buckling constraint and thermal load.

**Table 3:** Constraints and results for problem in Figure 10.

| Topology | Initial Constraints | Final Constraints | Volume & time (min) |
|---|---|---|---|
| Figure 11(a) | $J \leq 1.5 J_0$ $P \geq 0.4 P_0$ | $J = 1.44 J_0$ $\boxed{P = 0.4 P_0}$ | v=0.52 T=42 |
| Figure 11(b) | $J \leq 1.5 J_0$ $P \geq 0.4 P_0$ $\Delta t = 270^\circ C$ | $J = 1.17 J_0$ $\boxed{P = 0.4 P_0}$ $\Delta t = 270^\circ C$ | v=0.70 T=32 |

#### 4.2.1 wing-box optimization

Next, we consider optimization of the wing-box portion illustrated in Figure 12 where both the left and right ends are fixed, a lift pressure of $430$MPa is applied at the bottom, and a shear drag pressure of $146$MPa is exerted on the top edge. For FEA, 308,480 finite elements are used to discretize the design domain, leading to 1,022,328 DOF. The temperature rise is assumed to be $\Delta t = 170^\circ C$. The problem is posed as:

$$\underset{\Omega \subset D}{Min} |\Omega|$$
$$J \leq 3.5 J_0$$
$$P \geq 0.5 P_0 \qquad (57)$$
$$\text{subject to}$$
$$\mathbf{Kd} = \mathbf{f}_{st} + \mathbf{f}_{th}$$
$$\Delta t = 170^\circ C$$

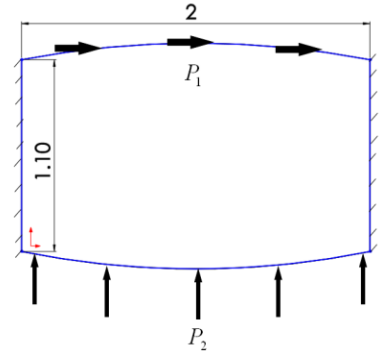

**Figure 12:** Rib wing-box portion with a thickness of 0.1 m, and applied boundary conditions.

The resulting topologies for the two scenarios (without and with thermal load) are shown in Figure 13. The results are summarized in Table 3. While both the buckling-compliance constrained problem (in Figure 13 (a)) and thermo-elastic optimization (in Figure 13 (b)) terminate due to buckling constraint, their optimized topologies have visibly different.

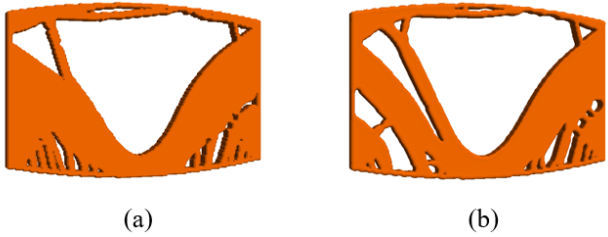

**Figure 13:** Optimal designs for the wing box portion: (a) With buckling constraint but no thermal load; (b) With buckling constraint and thermal load.

**Table 4:** Constraints and results for problem in Figure 13

| Topology | Initial Constraints | Final Constraints | Volume & time (min) |
|---|---|---|---|
| Figure 13(a) | $J \leq 3.5 J_0$  $P \geq 0.5 P_0$ | $J = 2.19 J_0$  $\boxed{P = 0.5 P_0}$ | v=0.53  T=63 |
| Figure 13(b) | $J \leq 3.5 J_0$  $P \geq 0.5 P_0$  $\Delta t = 170^\circ C$ | $J = 2.32 J_0$  $\boxed{P = 0.5 P_0}$  $\Delta t = 170^\circ C$ | v=0.51  T=71 |

It can be seen the optimized designs (in Figure 11 and Figure 13) are non-trivial and quite different from the traditional design (in Figure 9). By employing the proposed topology optimization method, the rib structure can be lightened by nearly 40% with a moderate compromise in stiffness and buckling resistance.

## 5. Conclusions

The main contribution of the paper is a new method for buckling constrained thermo-elastic topology optimization. Two different formulations were presented and compared. Both formulations exploit the concept of topological sensitivity; thus material parameterization is not required. As the numerical experiments reveal, the impact of thermal load on the final topologies can be significant for certain problems.

This paper is limited to linear buckling analysis and structure stiffness matrix is also assumed unchanged [57]. Although linear analysis is sufficient for simple thin plates and flat structures, non-linearity has to be considered in many situations where structures undergo significant pre-buckling rotations.

## Acknowledgements

The authors would like to thank the support of National Science Foundation through grants CMMI-1232508 and CMMI-1161474.